\documentclass{article}

\usepackage{times}
\usepackage{graphicx} 
\usepackage{subfigure} 

\usepackage{natbib}

\usepackage{algorithm}
\usepackage{algorithmic}

\usepackage{hyperref}

\usepackage[accepted]{icml2017}

\icmltitlerunning{Accelerating Green Computing with Hybrid Asymmetric Multicore  Architectures and Safe Parallelism}

\begin{document} 

\twocolumn[
\icmltitle{Accelerating Green Computing with Hybrid Asymmetric Multicore \\ Architectures and Safe Parallelism}



\icmlsetsymbol{equal}{*}

\begin{icmlauthorlist}
\icmlauthor{Hope Mogale}{equal,to}
\icmlauthor{Michael Esiefarienrhe}{equal,to,goo}
\icmlauthor{Naison Gasela}{goo}
\icmlauthor{Lucia Letlonkane}{ed}
\end{icmlauthorlist}

\icmlaffiliation{to}{North-West University, Mafikeng, South Africa}
\icmlaffiliation{goo}{North-West University, Mafikeng, South Africa}
\icmlaffiliation{ed}{North-West University, Mafikeng, South Africa}
\icmlaffiliation{to}{North-West University, Mafikeng, South Africa}

\icmlcorrespondingauthor{Hope Mogale}{Hope.Mogale@ieee.org}

\icmlkeywords{Multi-core architecture, Asymmetric cores, Eco-friendliness, Parallelism}

\vskip 0.3in
]



\printAffiliationsAndNotice{\icmlEqualContribution} 

\begin{abstract} 
  In this paper we present a novel strategy for accelarating green computing by utilizing and adopting the Hybrid Asymmetric Multicore Architectures (HAMA) model with Safe Parallelism.
  Most of the modern computing is serial and contributes to the global footprint of energy consumption. These impacts are often witnessed and experienced in
  many server farms and cloud computing platforms where the majority of the world's information resides. Evidently in this paper we present a novel strategy that
  can help decelerate the global footprint of energy consumption caused by computing. Through our strategy we prove that by adopting HAMA and
  utilizing safe parallelism energy consumption per computation can be minimized.
\end{abstract} 

\section{Introduction}
\label{submission}

Multicore microprocessors have been evolving since their inception in the early 2000s.
However, since their inception they have not been redesigned in a manner that is truly eco-friendly and energy efficient.
Microprocessors continue to be hazardous and none eco-friendly and the core of their problem lies in how they are manufactured as semiconductor devices. The current process of fabricating modern semiconductor devices is not energy efficient according to research in  \cite{gutowski2009thermodynamic}. Multicore architectures, when fully utilized, present us with great advantages in computation speeds as compared to the traditional monolithic processors. These parallel architectures can allow us to realize vast amounts of computing power enough to boost internet speeds, desired unlimited gaming, video streaming while at the same time multi tasking. The only problem is that programmers and application designers were slow in adopting a new style and approach to application development called parallel programming and \cite{pacheco2011introduction} hence the multicore technology and its underlying parallel architecture saw slow adoption rates from application developers. Programmers and application developers argue that parallel programming is not as mundane \cite{mccool2012structured, diaz2012survey} as traditional serial programming and development of applications. Using parallel programming as a style of development, developers would have to know well how to utilize parallel patterns and how objects compose, how to avoid races, and how to obtain deterministic output from code \cite{robison2013composable}. When programming for multicore architectures or any parallel architecture one has to know all boundaries and how to proof test that code is safe and obeys all laws of parallel programming such as the fundamental principle of computation which governs and dictates parallel computation known as Amdahl's Law \cite{amdahl1967validity}. Amdahl 's fundamental principle states that not all work can done in parallel and research \cite{hill2008amdahl} has proven that Amdahl 's Law still holds in the Multicore era.
There exists a taxonomy which represents classes of in which all computers can be classified and grouped according to their architectures. This was first illustrated and documented by Flynn \cite{flynn1972some} and has since become the de facto standard for classifying computers with their architectures \cite{flynn1972some} including the ones with parallel architectures. Further attempts have been proposed by researchers \cite{snyder1988taxonomy} to extend the taxonomy to accommodate other classes which may appear in future. Flynn's taxonomy identified that computers can be classified into four classes namely SIMD, SISD, MIMD, and MISD with variations between classes being the single data streams (SISD, MISD) and multiple data streams (SIMD, MIMD). Parallel computers and parallel architectures fall with the multiple data streams and these can be classified into two more categories of shared memory or distributed memory. When microprocessors adopted the multicore architectures scheme they fell into the multiple data stream. However not much has been developed to take advantage of their underlying architectures and research \cite{hill2008amdahl} shows that it is possible and much can still be done to improve utilization and reduce problems dark silicon \cite{esmaeilzadeh2011dark}. We aim to research and develop microprocessor architecture designs which are eco-friendly and suitable for promoting green computing. We will do all of this while our main emphasis remains energy efficiency. The technology treadmill continues to grow and elevate a staggering rate. As consumers continue to drive the technology treadmill the size of the transistors on microchips keep shrinking allowing devices such as wireless headsets to have embedded microchips. As this trend continue to grow problems arise in the semiconductor and the microprocessor industry. The amount of energy budget spent on today's microprocessors has reached unavoidable peaks. Transistors cannot shrink forever, and this suggests the end of Moore's law for Silicon. Researchers are now in search for a perfect candidate to replace silicon and this is referred to as the post-silicon era. The most important question is what will be that substitute.
All of that remains to be answered by time, while in the meantime the quest for research into developing better technology architectures
for multicore microprocessors continues. Our research presented in this paper is part of that quest. We shall structure the remainder of this paper
as follows. Section 2 will present research work related to our research then in section 3 will present our HAMA with Safe Parallelism, section 4 will discuss obtained experimental results. Lastly, section 5 will present conclusion and future work.

\section{Related Work} 
 
 We live in an age where every personal computer and other major technology products such as smart phones, smart watches and wireless
 earphones have microprocessor chips. Accompanying these chips are parallel architectures making high performance computing capabilities available for these microprocessor chips.
For over five decades \cite{mack2011fifty} these microprocessor chips have been governed by silicon and Moore's Law. This is now coming to an end and it is important to learn and investigate how green will the future be without silicon and its heavy ecological footprint. There has been a substantial amount of research published on determining how the future will be without silicon. In this section we identify, discuss and critique the most notable ones related to our research.
Massimo and team in \cite{fischetti2013theoretical} undertook a theoretical study to find out if silicon can withstand gate leakage currents below 10nm. Their brilliant theoretical study shows that scaling rules extracted from the 2011 International Technology Roadmap for Semiconductors (ITRS)-Roadmap and by using more strict scaling rules from the literature confirmed by simulations of 5-nm gate-length III VFETs. By employing local emperical pseudopotentials they were able see that gate current in the ON-state is shown to reach worrisome values at gate lengths of about 5 nm.
Kwon who is a senior researcher at the world's largest microchips producer samsung provided a notable research input in \cite{kwon2011eco}
which proposed a very important study that provides details on how to make the semiconductor industry power effiecient and how microchip can be produced in an
eco-friendly manner to protect the planet. In this study he outlined that to save mankind from life-threatening environmental crisis caused by non-ecofriendly
semiconductor technologies, the industry is expected to convert to eco-friendly technologies to preserve and save the environment. Kozawa et. al \cite{kozawa2014feasibility} conducted a study which investigated the extendibility of chemically amplified resist processes to the sub-10-nm half-pitch node taking into assumption the use of extreme ultraviolet lithography which demands more on the energy budget of a foundry. Furthermore, in this work they advise that although
sub-10-nm fabrication is considered to be feasible, a significant increase in the acid generation concentration and the development of related material technologies are required. Hence, this study declares jwwwthat as the shrinking continues the more energy will be needed to be consumed and therefore, more resources will have to be utilized which can have a negative outcome on the environment.
We aim to take all these studies into account when designing our hybrid asymmetric multicore architecture. We have used the lessons outlined by these studies to further enhance our architecture to make it more power efficient.

\section{HAMA with Safe Parallelism}
\label{submission}

\subsection{Hybrid Asymmetric Multicore Architecture (HAMA)}

In a similar body of work \cite{mogale2018introducing} previously proposed by same authors listed on this current research paper, a full description is given for a complete Hybrid Asymmetric Multicore Architecture called DOMINO. This paper is a continuation of research proposed by the authors in \cite{mogale2018introducing} and adopts the same definition for a Hybrid Asymmetric Multicore Architecture. As described by authors in \cite{mogale2018introducing} the Hybrid Asymmetric Multicore Architecture utilizes passive cores that are designed not to stay powered on at all times to preserve and save energy. Passive cores utilize a selftiming mechanism that is deterministic in nature and assures
that after completion of work the cores will be powered off. To avoid computation overheads these cores utilize structural parallelism which is fully compositional and highly deterministic by adopting parallel patterns.

\vspace{0.1in}

\centerline{
      \includegraphics[width=85mm]{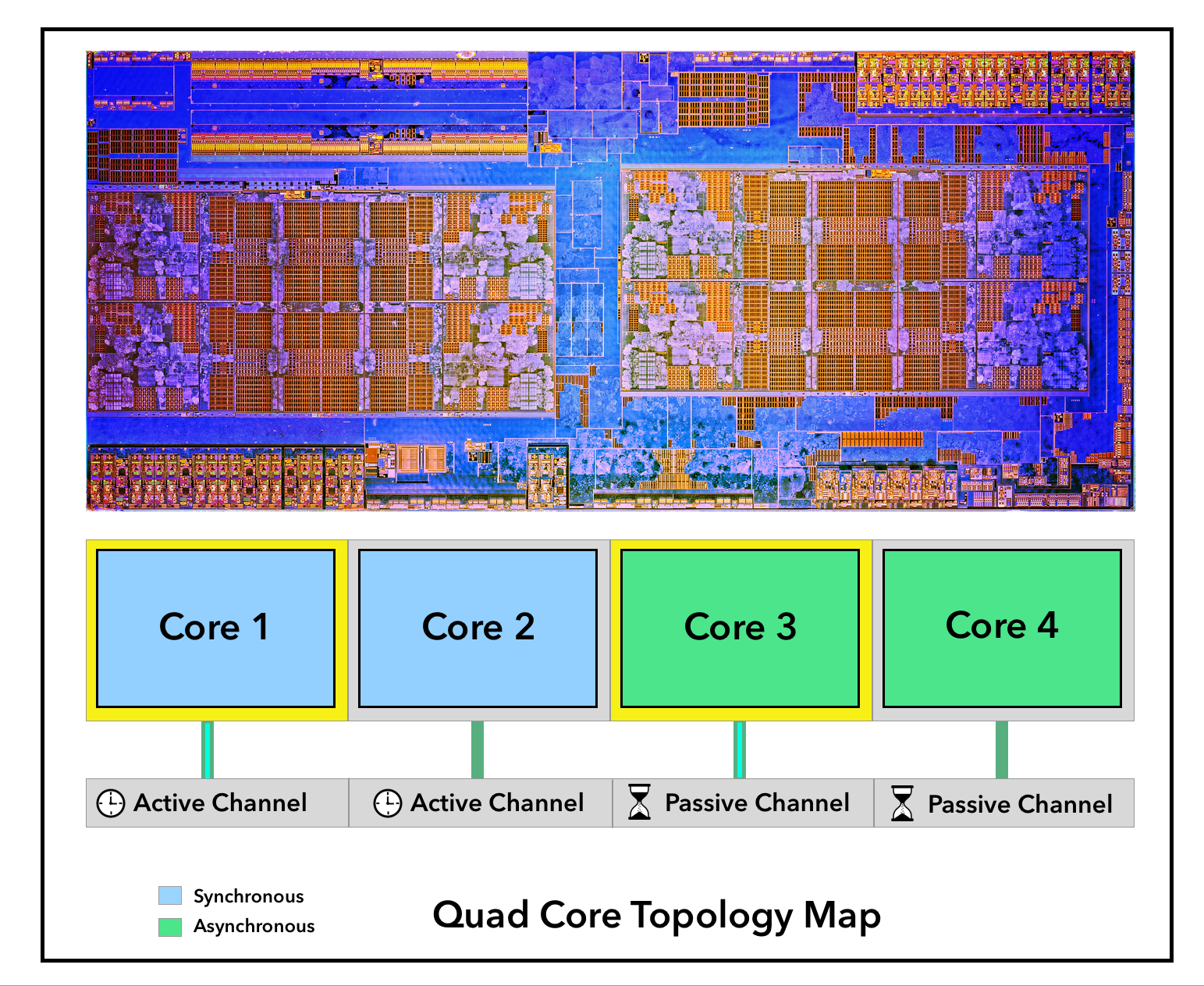}
    }

 \centerline{\bf Fig. 1. Quad Core Topology With Two Cores Powered.}

\vspace{0.1in}

  As seen in figure 1 workload is divided evenly on the passive cores using a designated parallel pattern as a guide since parallel patterns exhibit safe parallelism and ensure determinism.
Using parallel patterns such as a stencil or a map or a farm pattern we can synchronize the duration of the timer with that of the size of the problem in order to determine computation duration. We are aware that this entirely depends on Amdahl's law and we adopt asymmetric design coupled with asynchronous cores which can be passive to promote energy efficiency.

\subsection{Safe Parallelism with Parallel Patterns}

  For parallelism to be considered Safe it must be structurally defined with objects that compose and program code has to be fully deterministic without races, dependencies which can cause undesired computation overheads. Later in this paper we will demonstrate that parallelism is expensive in terms of energy consumption per computation for a typical multicore CPU. There exists several Algorthmic Skeleton Frameworks (AsKF) also known as parallel patterns which can be used to counter dark silicon and parallel computation problems \cite{mccool2012structured}, \cite{cole1989algorithmic} which occur frequently in Multicore Architecture Topologies. These Skeletons or parallel patterns seen in figure 2 are also known as
idoms in parallel computing, help maintain structered parallelism for fine grain computing \cite{robison2013composable}. Also they help make sure and maintain that computation does not only compose but also that it is deterministic at all times. Several of these skeletons exists in the serial programming world
but are insufficient for parallel computation. Most common known skeletons in the serial world are \texttt{for},
\texttt{if}, and \texttt{while}. These are not effective for parallel computation and so does not suit parallel architecures well
 and this has led researchers to develop new forms of Skeletons which can aid this problem. We discuss them briefly below:

\vspace{0.1in}
 \centerline{
       \includegraphics[width=80mm]{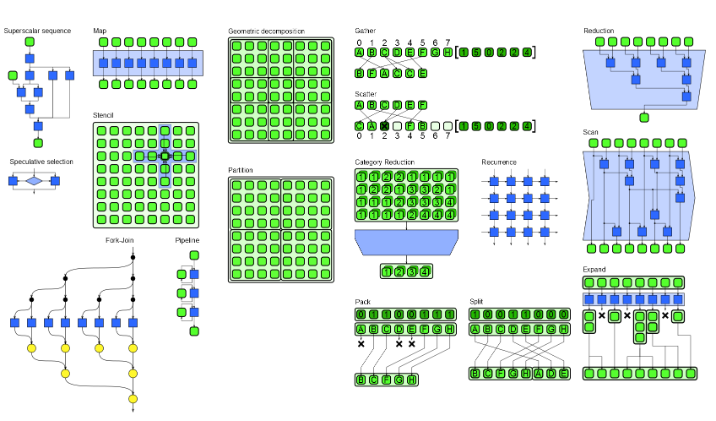}
     }
  \centerline{\bf Fig 2. Illustration of Parallel Patterns \cite{mccool2012structured}}
\vspace{0.1in}

\subsubsection{Map Pattern}

  The Map pattern is used for performing an operation on every element on a collection. The Map pattern
usually comes in handy in applications which utilize collections. On a map pattern, a serial iteration
pattern is executed which is independent by nature and has no dependencies. Illustrated in figure 3
the map pattern utilizes an elemental function which has known iterations to ensure composition and to eleminate
the problem of non-deterministic computation.  The map pattern is suitable for multicore architecures because
each strand of computation on each node in the map pattern sequence can be mapped to a core for parallel execution.

\vspace{0.1in}

\centerline{
      \includegraphics[width=50mm]{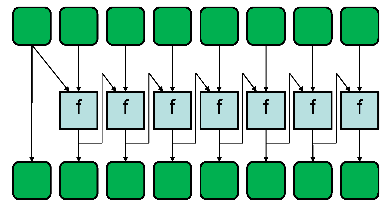}
    }
 \centerline{\bf Fig 3. Illustration of Map Pattern \cite{mccool2012structured}}
\vspace{0.1in}

  Map pattern is very suitable for embarrasingly parallel problems and can be nested with other patterns \cite{aldinucci2016parallel}, \cite{sheshikala2016parallel} to create a more powerful pattern for
computation. A map pattern can be advanced with a reduce pattern to form a Map-Reduce pattern which can help
enhance parallel computation. For example on a map-reduce pattern a mapper side $\big \langle x_1, y_1 \big \rangle $
together with its input $\big \langle x_2, y_2 \big \rangle $ shuffled and sorted and lastly given as input to
the reducer as $ \big \langle x_2, y_2 \big \rangle $ which then generates  $ \big \langle x_3, y_3 \big \rangle$ as
the last output. We discuss the reducer pattern next on our list.

\vspace{0.1in}

\subsubsection{Reduce Pattern}
  The reduce pattern is an idom which combines all items in a collection into one output. On a reduce pattern or a
reduction pattern which has a collection of $k$ items as an example, two adjacent items $x$ and $y$ of that
collection can be chosen and reduced to form a $k-1$ collection. Reduce pattern works well for applications such as
matrice multiplication and monte carlo simulation. It is used extensively in many algorithms because of its associative properties.
In summary it can be thought of as $ P = X_1 \oplus X_2 \oplus ... X_n $, provided $X_i$ represents the $ith$ item in the collection.
If we assume that a data collection $P$ is of type $c$ then we can use a binary function which is associative in nature. This function
can be represented as $ \oplus : c \times c \rightarrow c $, provided the function carries no dependencies.

\vspace{0.1in}
\subsubsection{Stencil Pattern}
A stencil pattern is a data pattern which behaves like a map pattern with the primary difference
being that the elemental function cannot only access the items in a collection but also items in a neibourhood.
Thus a stencil's output is a function of neighbourhoods of elements in a collection. A Stencil pattern is a variation
of the gather pattern and it is commonly used in applications of image processing which would normally utilize two dimensional
arrays for storing pixels for bitmaps and manipulating those pixels to achieve desired output as depicted in figure 4.

\vspace{0.1in}

\centerline{
      \includegraphics[width=70mm]{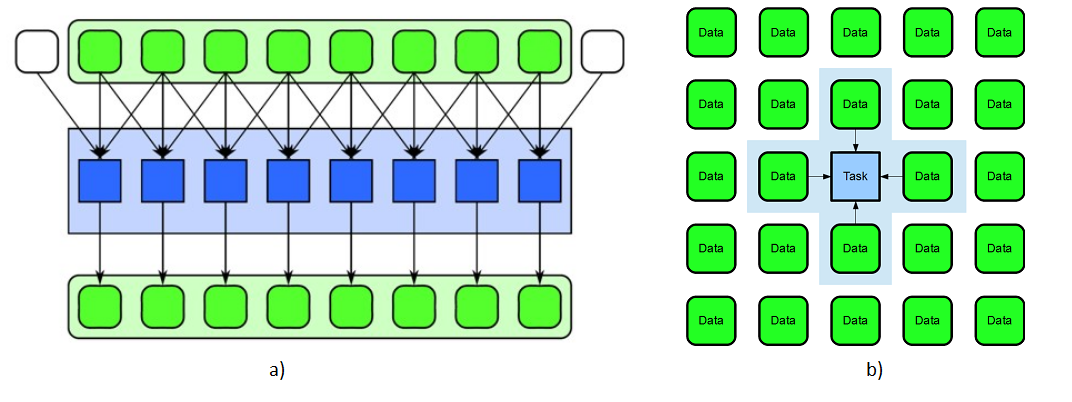}
    }
 \centerline{\bf Fig 4. Illustration of the Stencil Pattern \cite{mccool2012structured}}
\vspace{0.1in}

\subsubsection{Farm Pattern}
The farm pattern or parallel idom operates similar to the map pattern however, the size of the collection
is not known in advance. The farm pattern is also suitable for embarrasingly parallel computations.
One of the primary differences between the farm pattern and the map pattern is that the map pattern is a
data pattern while the farm pattern is a stream pattern.

\subsection{Karatsuba Polynomial Multiplication Experiment}

\subsubsection{Experimental Setup}

We have designed and optimized the Karatsuba Polynomial Multiplication algorithm which is used as a fast multiplication algorithm.
In our experiments this algorithm will be used to perform 256 multiplications over ten thousand degree Polynomials. Our target platform is both the old Haswell microarchitecture and the new Kaby Lake microarchitecture
all wrapped in a Core i7 microprocessor. The Kaby Lake die is of 14nm while Haswell is 22nm.
For both aforementioned nodes we have optimized the Karatsuba Algorithm described in Algorithm Listing 1 to be executed in serial, vectorized, Parallel, and
Parallel-Vectorized.
For parallel and Parallel vectorized we have optimized the algorithm to utilize the aforementioned parallel patterns as a strategy. We have timed both computations for
Haswell and Kaby Lake and after running the algorithm on the Haswell processor the following test results were obtained as seen in figure 5.

\vspace{0.1in}

\centerline{
      \includegraphics[width=85mm]{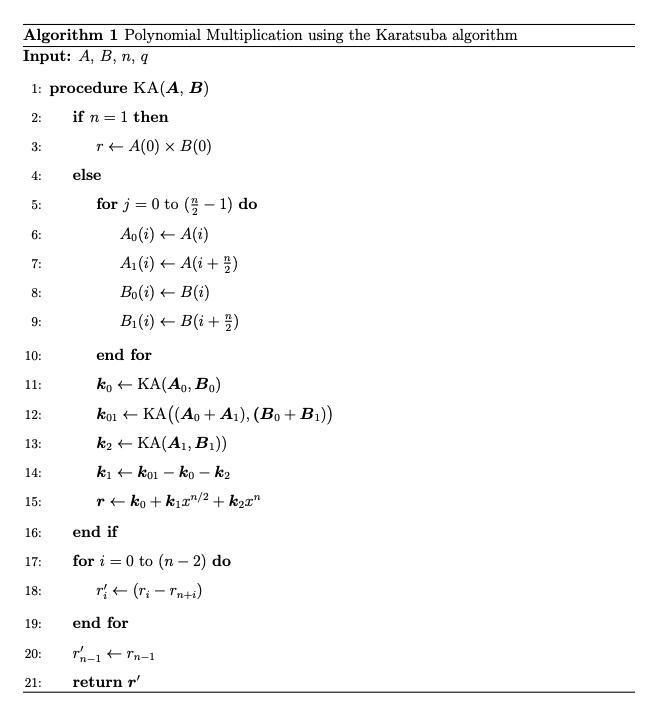}
    }
 \centerline{\bf Algorithm 1. Karatsuba Polynomial Multiplication}

\vspace{0.1in}

\subsubsection{Karatsuba on Haswell}

As can be seen from the screenshot in figure 5 the algorithm performed well based on how well it was optimized. As an example both parallel and vectorized parallel perform very well when compared to serial which in this case we can deem as suboptimal for speedup and performance. As we can see in figure 5 both parallel and Parallel/vectorized improved performance respectively with
3.85x and 3.97x speedup both with minimum timespan of 3.2 seconds over 12 seconds time span of both serial and vectorized.

\vspace{0.1in}

\centerline{
      \includegraphics[width=75mm]{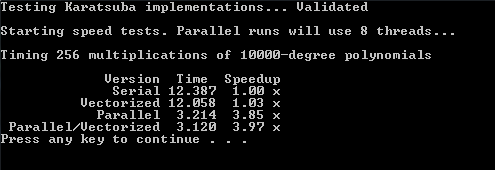}
    }
 \centerline{\bf Fig 5. Screenshot of Karatsuba on Haswell}
\vspace{0.1in}

  As visible in figure 5 our modified Karatsuba algorithm first runs serial, vectorized then follows parallel and Parallel/Vectorized.
If we look at the CPU utilization in terms of clockspeed we see that the algorithm utilizes the CPU. For both serial and vectorized the utilization
is not that heavy but for parallel and parallel-vectorized CPU Clock utilization is intense for all cores as visible in figure 6 below.

\vspace{0.1in}

\centerline{
      \includegraphics[width=85mm]{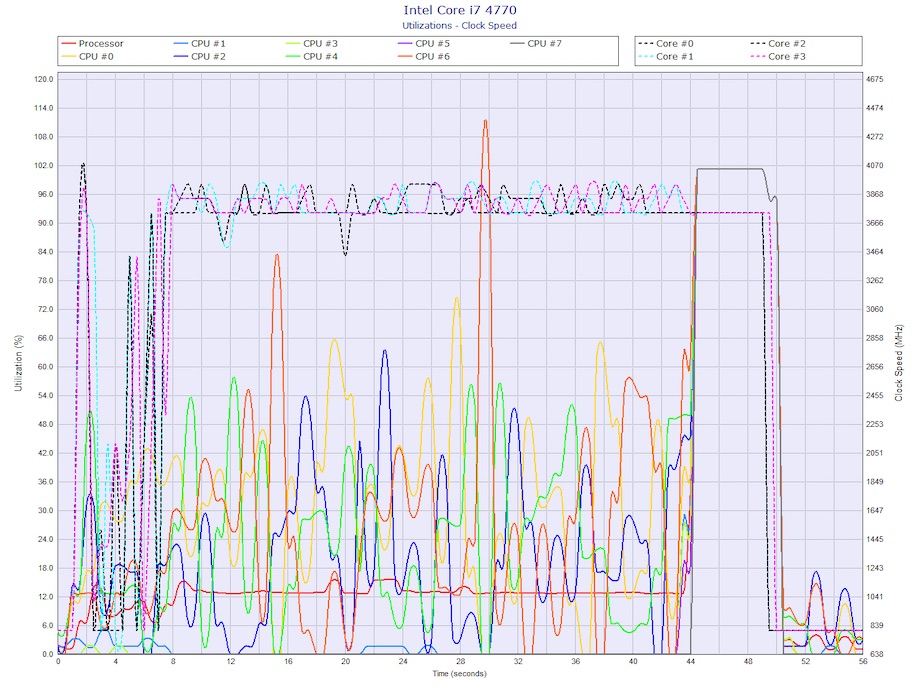}
    }
 \centerline{\bf Fig 6. CPU utilization of Karatsuba on Haswell}
\vspace{0.1in}

  For sake of performance, this is good. However it is wise to note that CPUs that are running at full speed consume a lot of energy. If we look at the
CPU temparature profile for the aforementioned experiment of Karatsuba on Haswell we notice that temparatures rise as soon as parallelism is utilized.
Hence this prompts us to come up with a better approach which promotes ecofriendly parallelism that is safe.

\vspace{0.1in}

\subsubsection{Karatsuba on Kaby Lake}
We have also tested Karatsuba running on Kaby Lake which is Intel's most recent state of the art microprocessor.
These microprocessors are designed to be efficient, powerful and since they are 14nm thick they feature fanless design.
We have our algorithm running on serial and optimized for vectorized, parallel and vector-parallel. The performance
we obtained can be seen visible in the screenshot of figure 7.

\vspace{0.2in}

\centerline{
      \includegraphics[width=75mm]{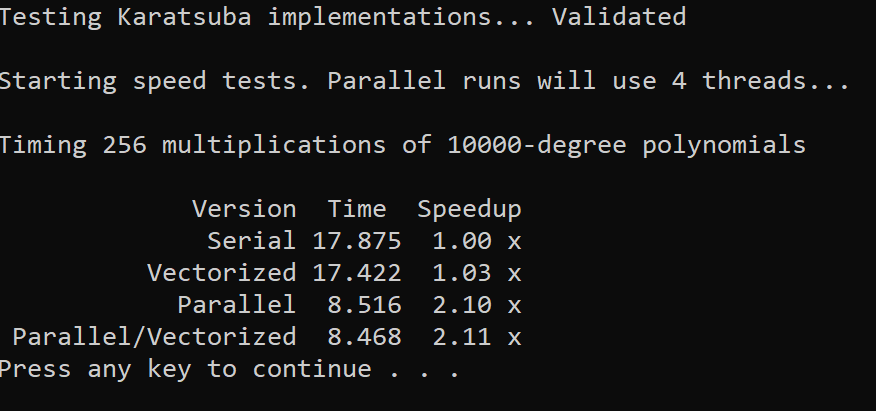}
    }
 \centerline{\bf Fig 7. Screenshot of Karatsuba on Kaby Lake}

\vspace{0.1in}

  We see from figure 7 that the results are quite different from Haswell which featured a full 8x Hyperthreaded
multicore microprocessor at 22nm. The serial part of the algorithm spanned 17.875s and a speedup of 1.00x was achieved
while the vectorized part performed at 17.422s with speedup of 1.03x which is better than the serial counterpart. Further,
we see that the parallel part of the algorithm halved the time by spanning only 8.516s to completed at achieved a speedup of 2.10 x.

\vspace{0.2in}

\centerline{
      \includegraphics[width=85mm]{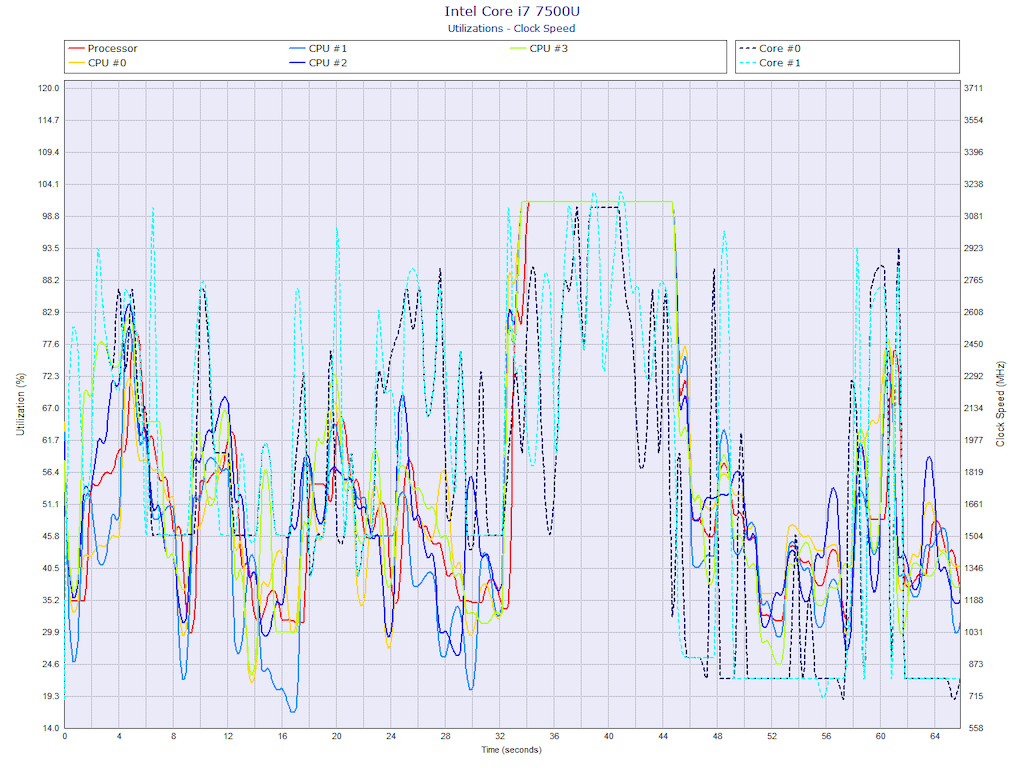}
    }
 \centerline{\bf Fig 8. CPU utilization of Karatsuba on Kaby Lake}

\vspace{0.1in}

  From figure 8 we can see that the portrayed clockspeed profile of Karatsuba running on Kaby Lake is totally different from the one of Haswell.
This is because at 14nm the core architecture has changed albeit with few similarities. One has to take into account the fact that the clockspeed of the
 processors is not the same, but the computation distribution pattern remains similar. We see that the last part which features vectorized parallelism is also flat. The reason why performance is great and spans the ideal optimal time for the vectorized algorithm is that our implementation combines both data parallelism by utilizing arrays and task parallelism by allowing compute intensive parts
of our algorithm such as polynomial computation to run in parallel. However as mentioned before, while parallelism reduces maximum computation time
it increases voltage utilization per core as visible in figure 8. This can be seen in figure 9 which presents CPU Voltage per core for Karatsuba running on Kaby Lake.

\vspace{0.2in}

\centerline{
      \includegraphics[width=80mm]{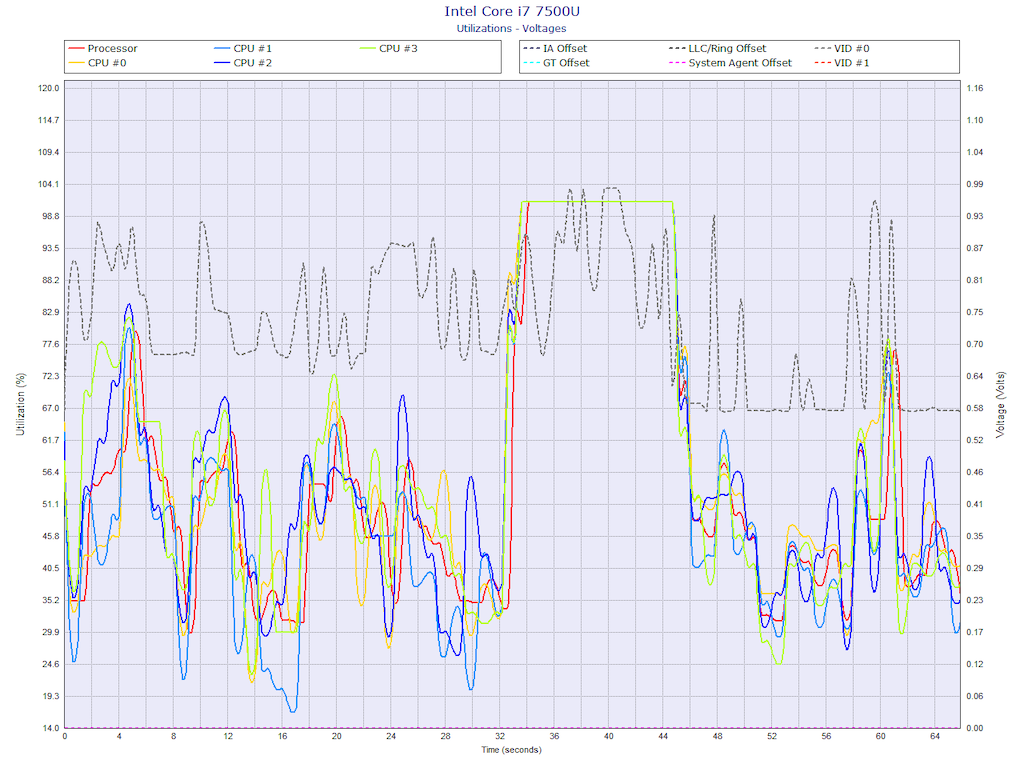}
    }
 \centerline{\bf Fig 9. Kaby Lake CPU Voltage per core}

\vspace{0.1in}

  If we look at both figures 8 and 12 we can notice few similarities. This is because the CPU Voltage performance
per core correlates with core clock speed. The higher the clockspeed the higher the Voltage per core and that is the drawback with multicore designs.
The clock speed is dependent on the overall thread activity that is experienced by the CPU. The total thread activity plays an important part in the performance of the CPU because if the workload is not balanced then performance deficiencies may be experienced. In this research we are promoting a strategy of using safe parallelism and Hybrid Asymmetric Multicores
as a strategy to accelerate green computing. We recognize that parallelism alone is not enough and that parallelism on multicore design is detrimental
because it increases energy use per core even though it reduces computation time as visible in figure 5 and figure 7.
To counter the energy use per core trap created by multicore designs we recommend adoption of our aforementioned hybrid assymmetric multicore designs which
feature low powered asynchronous cores. If we let all parallel activity to be run on these passive asynchronous cores then we believe that energy effiency can be
very much improved per computation since the asynchronous cores have no clock and feature low power coupled with globally asynchronous and locally synchronous strategy.
To provide brief evidence to this we provide a simplified energy analysis below.

\vspace{0.2in}

\subsubsection{Energy Effiency Analysis}

  We adopt Ahmdal's law collaries mentioned by authors in \cite{hill2008amdahl} for theoretical analysis and evaluation of our designs we take into consideration the following
theorem proposed by authors in \cite{yao2009extending} which states that if speedup for asymmetric is expressed as follows:

\vspace{0.2in}

\begin{center}
\(Speedup_{asymmetric} (f,n,r) = \frac{1}{(\frac{1-f}{perf(r)})+(\frac{f}{perf(r)+n-r})}\)                  (3.1)
\end{center}

\vspace{0.1in}

  it follows that if $ perf(r) = r^c, 0 <c< 1$, then it holds that:

\begin{itemize}
  \item If  \( \frac{f}{n-1} \frac{(1-c)}{c} \leq n^2 \) , then the maximum of speedup occurs at $r = 1$
  and the speedup is a decreasing function of $r$

  \item If \( \frac{f}{n} \geq \frac{n}{1-c} \) then it is clear that the maximum speedup occurs at $r=n$ and
  it is an increasing function of $r$

  \item Lastly, If  \( \frac{f}{n-1} \frac{(1-c)}{c} \leq n^2 \) and \( \frac{c}{f} \leq n^{1-c} \), then the maximum
  speedup will occur at a unique interval $r_0 \in (1, n)$.

\end{itemize}




\vspace{0.1in}

  Since we will be focusing on asymmetric design to improve perfomance we will take the aforementioned into consideration when analyzing the performance of our designs. However, one important aspect that we are very much concerned about is energy effieciency since the goal of this paper is to analyze and determine how to promote green computing by adopting HAMA and Safe Parallelism.

  If we take into account that energy effiency can be improved with parallelization, then we can deduce a strategic Axiom which states that:

\vspace{0.1in}

\begin{itemize}
  \item Processors can be run at an arbitrary clock frequency subject to a capped maximum frequency we will call $F_{max}$
  \item The speedup of $k$ that one can achieve with ideal Parallelism of $f = 0.99$ in correlation with processor speeds and scaling is
  subject to  \( 1 \leq k\leq \frac{1}{\frac{s+p}{N}} \) based on Amdahl's law defined in \cite{amdahl1967validity}.
  \item Lastly, we argue that albeit this is true we optimistically declare that the average computation span $k_{comp}$ approaches the ideal level of parallelization $f = 0.99$. That is the greater the value of $f$ gets, the less amount of $k_{comp}$ will be experienced and since $k_{comp}$ will be
  reduced, less energy will spent on an average computation.
\end{itemize}

\vspace{0.1in}

  It is worth noting that the last statement of our axiom takes into account the fact that parallelization increases both temparature and voltage per core peformance
as seen in figure 9 and figure 10 which presents the temparature profile of Karatsuba running on Haswell.

\vspace{0.2in}

\centerline{
      \includegraphics[width=85mm]{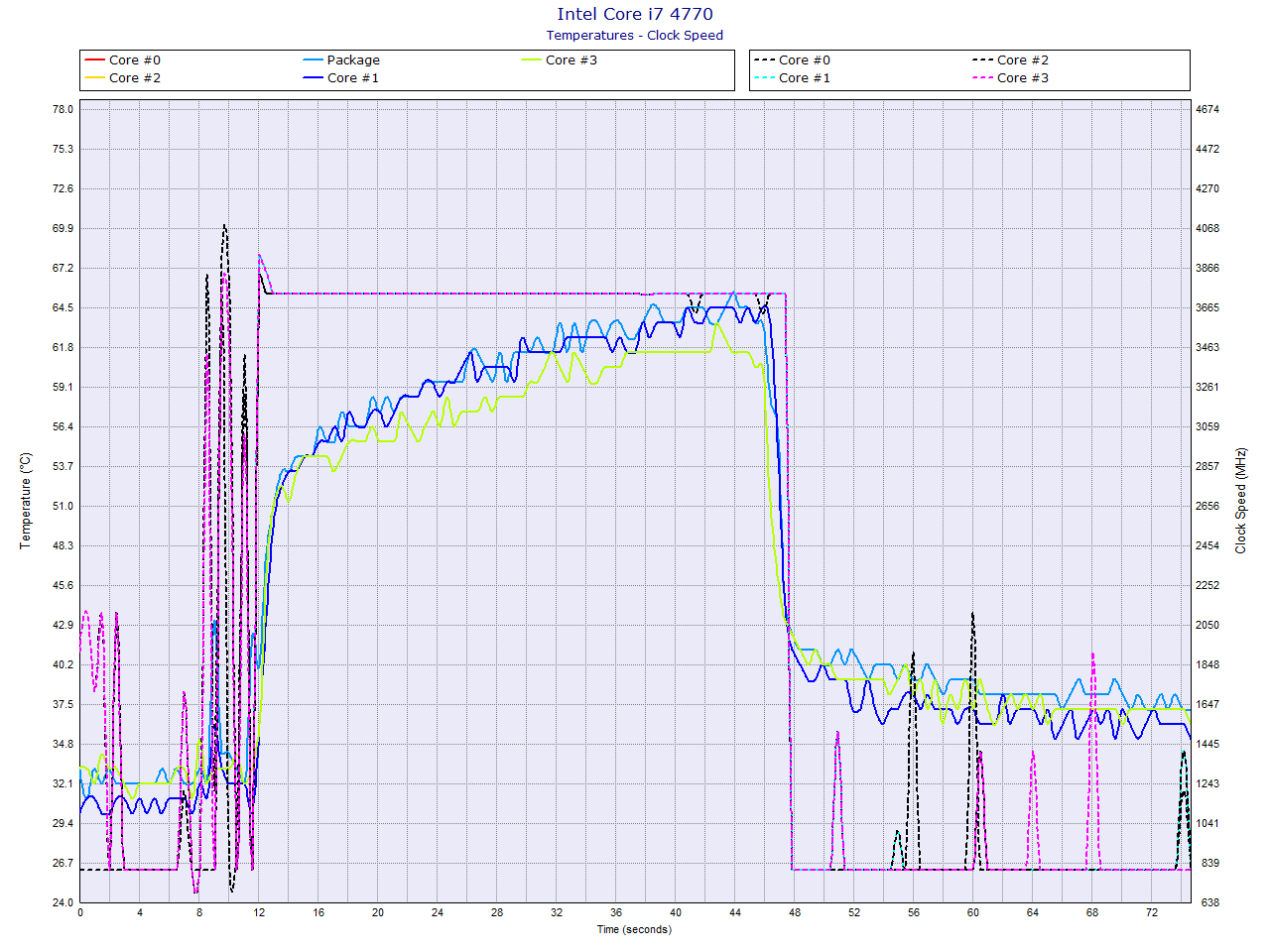}
    }
 \centerline{\bf Fig 10. Temparature Profile of Karatsuba on Haswell}

\vspace{0.1in}

\vspace{0.1in}
\subsection{Hybrid Architecture Perfomance Drawbacks}
  We would have liked to conclude this paper by telling researchers that there are no drawbacks to our designs.
 However, this can never be the case and the following is what we identified as drawbacks and we will be working
  on them in future research work. We also advise fellow researchers to lend a hand where possible since research is
a continous team effort.
\vspace{0.2cm}

\begin{itemize}
  \item \textbf{Core Computation Synchronization}  - Since our design features both synchronous and asynchronous cores on a die this gives birth to delay per computation due
  to synchronization. Even though this can be minimal at times it does become visible if workload is increased
  \item \textbf{Compatability Issues } - As of today most microprocessors utilize clocks and because operating systems have been programmed to use this clock, our design may not
  be compatible yet with many platforms becuase of the hybrid nature of the design causing performance to collapse since only synchronous cores may be recognized.
  \item \textbf{Sequential Synchronization} - Our design is truly optimized mostly for parallel computation hence the adoption of asynchronous cores. We believe that our design will
  experience under utilization if they are adopted for serial computation and since only synchronous active cores may be only active at a given time performance will be greatly reduced.

\end{itemize}

  To model power consumption for the last case we shall use the variable $D$ to refer to our asymmetric core. It is worth to note that $D$ is a special case that shall only arize when there is a need.
Using Amhdal's law collary described in equation 3.1, we know that $n$ is the number of processors and $f$ is the fraction of computation that can be parallelized $(0<=f<=1)$.
To model power consumption for $D$ we shall use $k$ as a variable for measuring the power consumed during idle time by our processor $(0<=k<=1)$.
Further, we assume that when our chip $D$ is in superscalar mode it consumes a power of 1. However, by definition using Amhdal's law \cite{amdahl1967validity}
the amount of power consumed by a processor during the sequential phase is 1 while the remaining $(n-1)$ processors consume $(n-1)k$. Thus we can assume that during
sequential computing phase our $D$ processor will consume is:

\vspace{0.1cm}
\begin{center}
\(Sequential = \frac{1 + (n-1)k}{(\frac{n}{2})}\)
\end{center}
\vspace{0.1cm}

  We note that this may not always be the case and as aforementioned $D$ is a special case for sequential computation which we did not particularly focus on as
part of our research goals.

\section{Conclusion and Future Work}
  In this paper we presented our novel strategy for accelerating green computing by utilizing and adopting Hybrid Asymmetric Multicore Architectures (HAMA)
model with Safe Parallelism. We have provided sufficient evidence which argues that parallelism alone is not enough to promote energy efficiency. Most of the modern computing is serial and contributes to the global footprint of energy consumption. These impacts are often witnessed and experienced in many server farms and cloud computing platforms where majority of the world's information resides.Through our novel strategy we have proven that by adopting HAMA and utilizing safe parallelism, energy consumption per computation can be greatly minimized. We have also outlined the minor drawbacks of our strategy which we hope overcome with future work.

\bibliography{example_paper}

\begin{thebibliography}{19}
\providecommand{\natexlab}[1]{#1}
\providecommand{\url}[1]{\texttt{#1}}
\expandafter\ifx\csname urlstyle\endcsname\relax
  \providecommand{\doi}[1]{doi: #1}\else
  \providecommand{\doi}{doi: \begingroup \urlstyle{rm}\Url}\fi

\bibitem[Aldinucci et~al.(2016)Aldinucci, Danelutto, Drocco, Kilpatrick,
  Misale, Pezzi, and Torquati]{aldinucci2016parallel}
Aldinucci, Marco, Danelutto, Marco, Drocco, Maurizio, Kilpatrick, Peter,
  Misale, Claudia, Pezzi, G~Peretti, and Torquati, Massimo.
\newblock A parallel pattern for iterative stencil+ reduce.
\newblock \emph{The Journal of Supercomputing}, pp.\  1--16, 2016.

\bibitem[Amdahl(1967)]{amdahl1967validity}
Amdahl, Gene~M.
\newblock Validity of the single processor approach to achieving large scale
  computing capabilities.
\newblock In \emph{Proceedings of the April 18-20, 1967, spring joint computer
  conference}, pp.\  483--485. ACM, 1967.

\bibitem[Cole(1989)]{cole1989algorithmic}
Cole, Murray~I.
\newblock \emph{Algorithmic skeletons: structured management of parallel
  computation}.
\newblock Pitman London, 1989.

\bibitem[Diaz et~al.(2012)Diaz, Munoz-Caro, and Nino]{diaz2012survey}
Diaz, Javier, Munoz-Caro, Camelia, and Nino, Alfonso.
\newblock A survey of parallel programming models and tools in the multi and
  many-core era.
\newblock \emph{Parallel and Distributed Systems, IEEE Transactions on},
  23\penalty0 (8):\penalty0 1369--1386, 2012.

\bibitem[Esmaeilzadeh et~al.(2011)Esmaeilzadeh, Blem, Amant, Sankaralingam, and
  Burger]{esmaeilzadeh2011dark}
Esmaeilzadeh, Hadi, Blem, Emily, Amant, Renee~St, Sankaralingam, Karthikeyan,
  and Burger, Doug.
\newblock Dark silicon and the end of multicore scaling.
\newblock In \emph{Computer Architecture (ISCA), 2011 38th Annual International
  Symposium on}, pp.\  365--376. IEEE, 2011.

\bibitem[Fischetti et~al.(2013)Fischetti, Fu, and
  Vandenberghe]{fischetti2013theoretical}
Fischetti, Massimo~V, Fu, Bo, and Vandenberghe, William~G.
\newblock Theoretical study of the gate leakage current in sub-10-nm
  field-effect transistors.
\newblock \emph{IEEE Transactions on Electron Devices}, 60\penalty0
  (11):\penalty0 3862--3869, 2013.

\bibitem[Flynn(1972)]{flynn1972some}
Flynn, Michael~J.
\newblock Some computer organizations and their effectiveness.
\newblock \emph{IEEE transactions on computers}, 100\penalty0 (9):\penalty0
  948--960, 1972.

\bibitem[Gutowski et~al.(2009)Gutowski, Branham, Dahmus, Jones, Thiriez, and
  Sekulic]{gutowski2009thermodynamic}
Gutowski, Timothy~G, Branham, Matthew~S, Dahmus, Jeffrey~B, Jones, Alissa~J,
  Thiriez, Alexandre, and Sekulic, Dusan~P.
\newblock Thermodynamic analysis of resources used in manufacturing processes.
\newblock \emph{Environmental science \& technology}, 43\penalty0 (5):\penalty0
  1584--1590, 2009.

\bibitem[Hill \& Marty(2008)Hill and Marty]{hill2008amdahl}
Hill, Mark~D and Marty, Michael~R.
\newblock Amdahl's law in the multicore era.
\newblock 2008.

\bibitem[Kozawa et~al.(2014)Kozawa, Santillan, and
  Itani]{kozawa2014feasibility}
Kozawa, Takahiro, Santillan, Julius~Joseph, and Itani, Toshiro.
\newblock Feasibility study of sub-10-nm half-pitch fabrication by chemically
  amplified resist processes of extreme ultraviolet lithography: I. latent
  image quality predicted by probability density model.
\newblock \emph{Japanese Journal of Applied Physics}, 53\penalty0
  (10):\penalty0 106501, 2014.

\bibitem[Kwon(2011)]{kwon2011eco}
Kwon, Oh-Hyun.
\newblock Eco-friendly semiconductor technologies for healthy living.
\newblock In \emph{Solid-State Circuits Conference Digest of Technical Papers
  (ISSCC), 2011 IEEE International}, pp.\  22--28. IEEE, 2011.

\bibitem[Mack(2011)]{mack2011fifty}
Mack, Chris~A.
\newblock Fifty years of moore's law.
\newblock \emph{IEEE Transactions on semiconductor manufacturing}, 24\penalty0
  (2):\penalty0 202--207, 2011.

\bibitem[McCool et~al.(2012)McCool, Reinders, and
  Robison]{mccool2012structured}
McCool, Michael, Reinders, James, and Robison, Arch.
\newblock \emph{Structured parallel programming: patterns for efficient
  computation}.
\newblock Elsevier, 2012.

\bibitem[Mogale et~al.(2018)Mogale, Esiefarienrhe, Gasela, and
  Letlonkane]{mogale2018introducing}
Mogale, Hope, Esiefarienrhe, Michael, Gasela, Naison, and Letlonkane, Lucia.
\newblock Introducing domino: An eco-friendly asynchronous hybrid multicore
  architecture for green computing.
\newblock In \emph{2018 International Conference on Advances in Big Data,
  Computing and Data Communication Systems (icABCD)}, pp.\  1--7. IEEE, 2018.

\bibitem[Pacheco(2011)]{pacheco2011introduction}
Pacheco, Peter.
\newblock \emph{An introduction to parallel programming}.
\newblock Elsevier, 2011.

\bibitem[Robison(2013)]{robison2013composable}
Robison, Arch~D.
\newblock Composable parallel patterns with intel cilk plus.
\newblock \emph{Computing in Science \& Engineering}, 15\penalty0 (2):\penalty0
  0066--71, 2013.

\bibitem[Sheshikala et~al.(2016)Sheshikala, Rao, and
  Prakash]{sheshikala2016parallel}
Sheshikala, M, Rao, D~Rajeswara, and Prakash, R~Vijaya.
\newblock Parallel approach for finding co-location pattern--a map reduce
  framework.
\newblock \emph{Procedia Computer Science}, 89:\penalty0 341--348, 2016.

\bibitem[Snyder(1988)]{snyder1988taxonomy}
Snyder, Lawrence.
\newblock A taxonomy of synchronous parallel machines.
\newblock Technical report, DTIC Document, 1988.

\bibitem[Yao et~al.(2009)Yao, Bao, Tan, and Chen]{yao2009extending}
Yao, Erlin, Bao, Yungang, Tan, Guangming, and Chen, Mingyu.
\newblock Extending amdahl's law in the multicore era.
\newblock \emph{ACM SIGMETRICS Performance Evaluation Review}, 37\penalty0
  (2):\penalty0 24--26, 2009.

\end{thebibliography}
\bibliographystyle{icml2017}

\end{document}